\documentclass[a4paper,11pt]{article}
\usepackage{pos}
\usepackage[normalem]{ulem}

\title{Inclusive $\tau$ Hadronic Decay Rate in a Renormalon-Free Gluon Condensate Scheme}

\author[a]{Miguel A. Benitez-Rathgeb}
\author[a,b]{Diogo Boito}
\author*[a,c]{Andr\'e H. Hoang}
\author[a,d]{Matthias Jamin}
\author[a,e]{Christoph Regner}

\affiliation[a]{University of Vienna, Faculty of Physics, Boltzmanngasse 5, A-1090 Wien, Austria}

\affiliation[b]{Instituto de F\'isica de São Carlos, Universidade de S\~ao Paulo, CP 369, 13560-970, S\~ao Carlos, SP, Brazil}

\affiliation[c]{Erwin Schr\"odinger International Institute for Mathematics and Physics, University of Vienna, Boltzmanngasse 9, A-1090 Wien, Austria}

\affiliation[d]{Department of Addictive Behaviour, Central Institute of Mental Health, Medical Faculty Mannheim, Heidelberg University, Mannheim, Germany}

\affiliation[e]{University of Vienna, Vienna Doctoral School in Physics, Boltzmanngasse 5, A-1090 Wien, Austria}

\emailAdd{miguel.angel.benitez-rathgeb@univie.ac.at}
\emailAdd{boito@ifsc.usp.br}
\emailAdd{andre.hoang@univie.ac.at}
\emailAdd{matthias.jamin@gmail.com}
\emailAdd{christoph.regner@univie.ac.at}

\abstract{In a recent work by some of us it was shown that the long-standing discrepancy between the QCD perturbation series for the inclusive hadronic tau decay rate computed in the CIPT and FOPT expansion approaches can be understood from the fact that CIPT has an infrared (IR) sensitivity that it not compatible with the standard form of the operator production expansion (OPE). For concrete IR renormalon models for the QCD Adler function the resulting CIPT-FOPT discrepancy, the asymptotic separation, can be calculated analytically from the Borel representation of the CIPT series expansion. 
If the known perturbative corrections for the QCD Adler function at the 5-loop level already have a sizeable contribution from the asymptotic behavior related to the gluon condensate (GC) renormalon, the asymptotic separation is dominated by that renormalon. This implies that the CIPT expansion can be reconciled with FOPT, when 
a renormalon-free scheme for the GC is adopted. In this talk we discuss such a renormalon-free scheme for the GC, which involves perturbative subtractions in analogy to using short-distance quark mass schemes instead of the pole mass.
Using a concrete realistic high-order Borel model for the Adler function consistent with the known corrections up to 5 loops and containing a sizeable GC renormalon contribution, we show that the CIPT-FOPT discrepancy can be avoided when switching to the renormalon-free GC scheme. At the same time, the perturbative convergence of $\tau$ hadronic spectral funtion moments strongly sensitive to the GC OPE corrections is considerably improved. We show that these improvements may lead to higher precision for strong coupling determinations.
}

\FullConference{%
  Loops and Legs in Quantum Field Theory - LL2022,\\
  25-30 April, 2022\\
  Ettal, Germany
}

\begin{document}
\maketitle

\section{Introduction}

Extractions of the strong coupling $\alpha_s$ from inclusive hadronic $\tau$ decay spectral function moments represents one of the most precise methods to obtain this fundamental parameter of QCD from experimental data. The 
determinations are based on weighted finite-energy sum rule integrals (spectral function moments) over the experimental inclusive $\tau$ hadronic spectral functions compared to the corresponding theory predictions. The inclusive $\tau$ hadronic decay width 
$R_{\tau} \equiv \frac{\Gamma[\tau^{-} \to {\rm hadrons} ~ \nu_{\tau}(\gamma)]}{\Gamma[\tau^{-} \to e^{-} \overline{\nu}_e\nu_{\tau}(\gamma)]}$
is a particular case of these spectral function moments. In the chiral limit the theoretical spectral function moments can be dissected into the following individual contributions,
\begin{align}
\label{eq: theory dissection}
R^{(W)}(s_0) \, =\, \frac{3}{2} \,S_{\rm ew}\,|V_{ud}|^2 \Big[\,
\delta^{\rm tree}_{W} + \delta^{(0)}_{W}(s_0)  +
\sum_{d\geq 4}\delta^{(d)}_{W}(s_0) +\delta_{W}^{(\rm DV)}(s_0)\Big] \,,
\end{align} 
where $S_{\rm ew}$ denotes electroweak corrections and $s_0$ is the upper bound of the weighted spectral function integrations. $\delta^{\rm tree}_{W}$ corresponds to the tree level contribution while  $\delta^{(0)}_{W}(s_0)$ represents the perturbative QCD corrections. The index $W$ refers to the weight function $W(x)$ which satisfies $W(1)=0$ and typically is a polynomial in $x$. For the inclusive decay rate, which is called the kinematic moment, this polynomial is $W_{\rm kin}=(1-x)^3(1+x)=1-2x+2x^3-x^4$ and we have $s_0=m_\tau^2$. The term $\delta^{(d)}_{W}(s_0)$ denotes non-perturbative higher dimensional corrections in the operator product expansion (OPE). The last term in Eq.~(\ref{eq: theory dissection}), $\delta_{W}^{(\rm DV)}(s_0)$, stands for the duality violation (DV) contributions, which are suppressed for weight functions with $\frac{d}{dx}W(1)=0$ and not further considered throughout this talk.

The QCD corrections are given by contour integrals of the form ($x \equiv \frac{s}{s_0}$)
\begin{equation}
\label{eq:deltadef}
\delta^{(0)}_{W} (s_0)\, =\, \frac{1}{2\pi i}\,\,\ointctrclockwise_{|s|=s_0}\!\! \frac{{\rm d}s}{s}\,W ({\textstyle \frac{s}{s_0}})\,\hat D(s) ,
\end{equation}
involving the (reduced) partonic Adler function $\hat{D}(s)$ for invariant mass squared $s$,
\begin{eqnarray}
	\label{eq:AdlerseriesCIPT}
	\hat D(s) & \, =  \,& \, \sum\limits_{n=1}^\infty
	c_{n,1} \,\big({\textstyle \frac{\alpha_s(-s)}{\pi}}\big)^n\,, 
	\\ \label{eq:AdlerseriesFOPT}
	& \, = \, &
	\, \sum\limits_{n=1}^\infty\,
	\big({\textstyle \frac{\alpha_s(s_0)}{\pi}}\big)^n \, \sum\limits_{k=1}^{n} k\, c_{n,k}\,\ln^{k-1}({\textstyle \frac{-s}{s_0}}) \,.
\end{eqnarray}
The contour integral can be analytically related to an integration along the real positive $s$-axis with $s_0$ being the upper bound.
The conventional integration path along the circle with $|s| = s_0$ may be deformed as long as it remains in the perturbative region. 
The nonperturbative terms $\delta^{(d)}_{W}(s_0)$ can be obtained from an analogous integral over the Adler function's OPE corrections ($a(\mu^2)=\alpha_s(\mu^2)\beta_0/4\pi$)
\begin{align}
\label{eq:OPEstandard}
D^{\rm OPE}(s) \,=\, 
\frac{2\pi^2}{3}\,\frac{1 -\frac{22}{81} \, a(-s) + \ldots}{s^2}\, \langle \bar G^2\rangle 
+\sum_{d=6}^\infty \frac{1}{(-s)^{d/2}} \sum_i  C_{d,i}(\alpha_s(-s)) \langle \bar {\cal O}_{d,\gamma_i}\rangle \,,
\end{align}
where the term that is leading in the power expansion in $\Lambda_{\rm QCD}^2/s$ involves the renormalization scheme-invariant GC matrix element $\langle \bar G^2\rangle=\langle\Omega| (\frac{\alpha_s}{\pi}+\ldots)G^{\mu\nu}G_{\mu\nu}|\Omega \rangle$. For spectral function moments with a weight function without a quadratic $x^2$ term, such as $W_{\rm kin}$, the GC correction is very strongly suppressed (and even vanishing when the higher order corrections in its Wilson coefficient are neglected).

In the theory integrals a prescription for setting the renormalization scale $\mu$ has to be adopted and the two most commonly considered prescriptions are known as fixed-order (FOPT), where $\delta^{(0)}_{W} (s_0)$ is an expansion in powers of $\alpha_s(s_0)$ (see Eq.~(\ref{eq:AdlerseriesFOPT})), and contour-improved perturbation theory (CIPT)~\cite{Pivovarov:1991rh}, where the expansion is in powers of $\alpha_s(-s)$ before doing the contour integration (see Eq.~(\ref{eq:AdlerseriesCIPT})). When determining the strong coupling from $\tau$ hadronic spectral function moments, the discrepancy of the results obtained from using either FOPT or CIPT has been discussed controversially~\cite{Descotes-Genon:2010pyp,Beneke:2012vb} and up to now constituted one of the dominant uncertainties in the determination of $\alpha_s$ from hadronic $\tau$ decays. In general, CIPT leads to smaller values for $\delta^{(0)}_{W} (s_0)$ than FOPT, which typically results in larger extracted values for $\alpha_s (m_\tau^2)$ when the CIPT expansion is employed.

\section{Asymptotic Separation}

In a recent work two of us have provided a conceptual as well as quantitative analysis of the observed CIPT-FOPT discrepancy. In contrast to previous studies~\cite{Beneke:2008ad,Descotes-Genon:2010pyp,Beneke:2012vb}, where it was assumed that the Borel representations for the FOPT and CIPT expansion series are identical, it was shown in Ref.~\cite{Hoang:2020mkw,Hoang:2021nlz} that their Borel representations are in fact intrinsically different in the presence of IR renormalons. In particular, the expressions for the two non-equivalent representations are given by
\begin{align}
\label{eq:BorelFOPT} 
 \delta_{W,{\rm Borel}}^{(0),{\rm FOPT}}(s_0) &= {\rm PV} \int_0^\infty \!\! {\rm d} u \,\, 
\frac{1}{2\pi i}\,\ointctrclockwise_{|x|=1} \frac{{\rm d}x}{x} \, W(x) \,
B[\hat D](u)\,e^{-\frac{u}{a(-xs_0)}}\,, 
\end{align}
\vspace{-0.5cm}
\begin{align}
\label{eq:BorelCIPT} 
 \delta_{W,{\rm Borel}}^{(0),{\rm CIPT}}(s_0) &= \int_0^\infty \!\! {\rm d} \bar u \,\,  
\frac{1}{2\pi i}\,\ointctrclockwise_{{\cal C}_x} \frac{{\rm d}x}{x} \, W(x) \,
\big({\textstyle \frac{\alpha_s(-x s_0)}{\alpha_s (s_0)}}\big)\,
B[\hat D]\Big({\textstyle \frac{\alpha_s(-x s_0)}{\alpha_s (s_0)}} \bar u\Big)
\,e^{- \frac{\bar u}{a(s_0)}}\,, 
\end{align}
where $B[\hat{D}](u)$ is the Borel function of the real-valued Euclidean Adler function $\hat{D}(-s_0)$ with respect to an expansion in $\alpha_s(s_0)$, in the sense that the $u$-Taylor series of $B[\hat{D}](u)$ gives the series in powers of $\alpha_s(s_0)$ through the Borel integral
\begin{align}
\hat{D} (-s_0) = \int^\infty_0 \mathrm{d} u \,  B[\hat{D}] (u) \,  \mathrm{e}^{-\frac{ u}{a (s_0)}} \, .
\end{align} 
The previously known FOPT Borel representation in Eq. (\ref{eq:BorelFOPT}) includes the commonly employed principal value  (PV) prescription to obtain a well-defined result for the value of the Borel representation (briefly called the Borel sum). This is due to the IR renormalon cuts (or poles) contained in $B[\hat D](u)$ along the positive real $u$-axis.
The novel Borel representation for the CIPT series in Eq.~(\ref{eq:BorelCIPT}) implies a different regularization of the non-analytic IR renormalon cuts because $\frac{\alpha_s(-x s_0)}{\alpha_s (s_0)} \bar u$ is complex-valued. Due to the fact that the complex-valued coupling $\alpha_s (-x s_0)$ is now part of the argument of the Borel function $B [\hat{D}]$, the path ${\cal{C}}_{x}$ needs to be deformed away from the circle $|x|=1$ to account for the modified singularity structure when the CIPT Borel sum is evaluated. 

\begin{figure}
\begin{center}
\includegraphics[width=0.75\textwidth,angle=0]{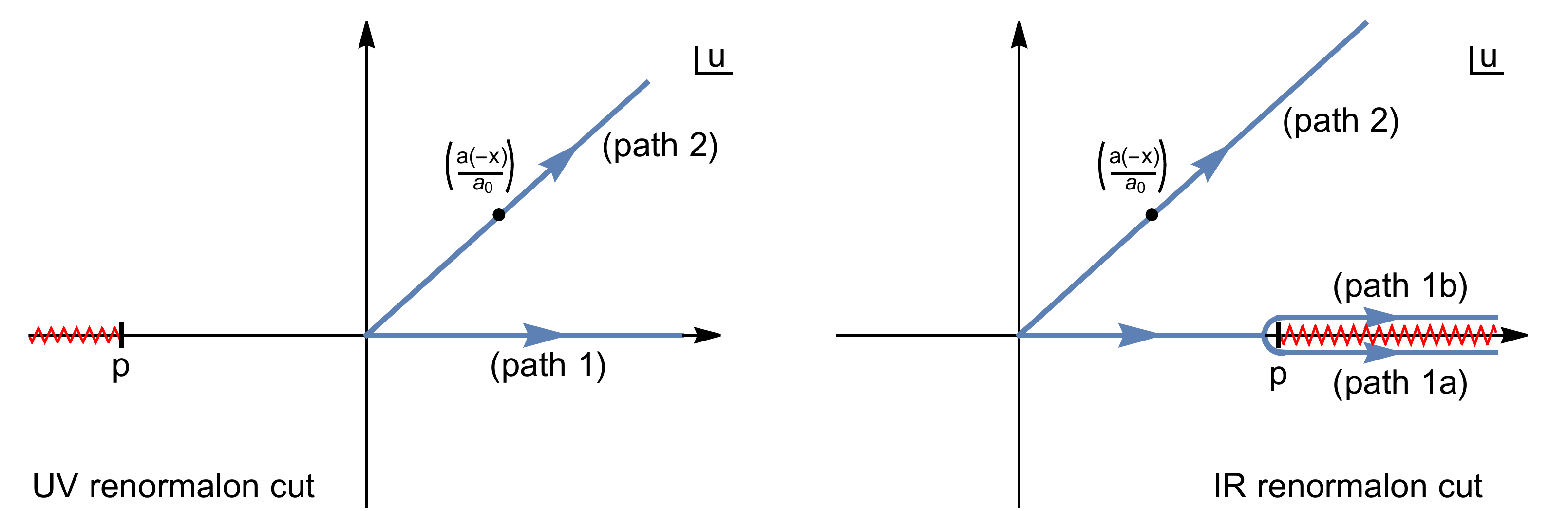}
\caption{\label{fig:contour}
Graphical representation of the integration paths in the complex Borel plane involved for the FOPT and CIPT approach for the cases of a UV renormalon ($p<0$, left panel) and an IR renormalon ($p>0$, right panel). The red zig-zag lines denote renormalon cuts starting at $u=p$. Figures taken from Ref.~\cite{Hoang:2020mkw}.}
\end{center}
\end{figure}

\begin{figure}
\begin{center}
\includegraphics[width=\textwidth]{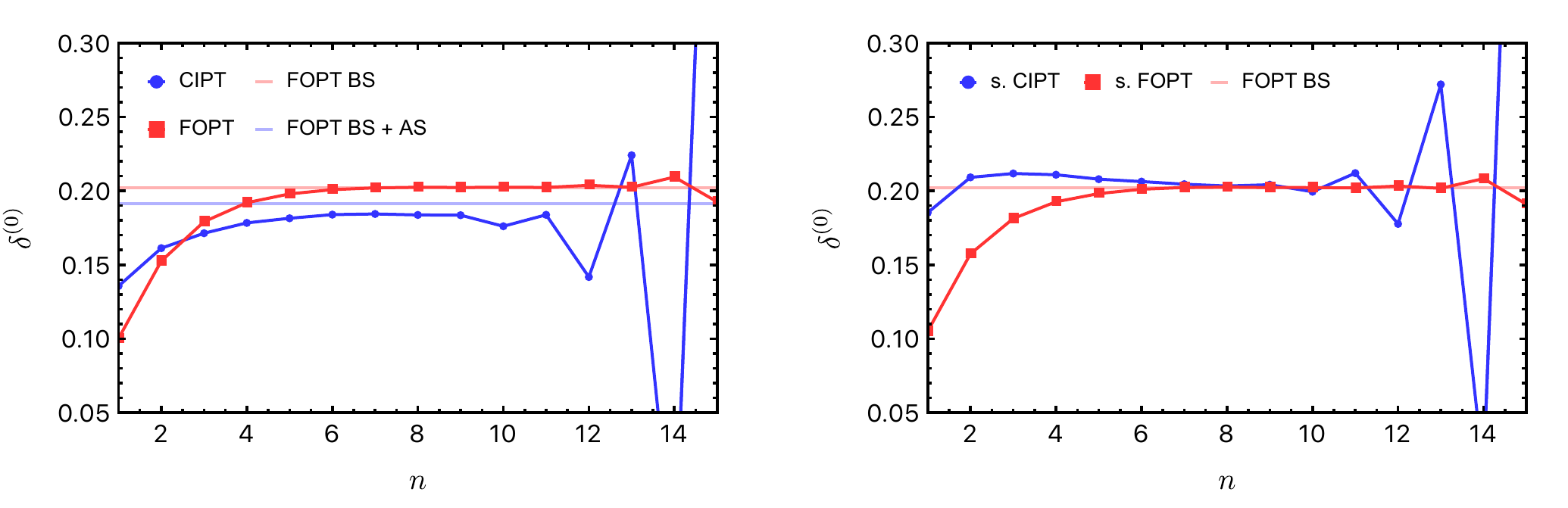}
\caption{\label{Plot-2}
Perturbative series for the kinematic moment for $s_0=m_\tau^2$ in the FOPT and CIPT expansions as a function of order $n$ in the $\overline{\rm MS}$ GC scheme
(left panel) and the RF GC scheme for $R=0.8\,m_\tau$ and $N_g=0.64$ (right panel) using the Borel model in section~6 of Ref. \cite{Beneke:2008ad} for $n>5$. The red and blue horizontal lines represents the Borel sums (BSs) of the FOPT and CIPT series, respectively.}
\end{center}
\end{figure}

The Borel representations are formally related through the complex-valued change of variables $u= \bar{u}\,  \alpha_s (-x s_0) / \alpha_s (s_0)$, which (as illustrated in Fig.~\ref{fig:contour} for ${\rm Im}[\alpha_s(-xs_0)]>0$) leads to a difference between $\delta_{W,{\rm Borel}}^{(0),{\rm FOPT}} (s_0)$ and $\delta_{W,{\rm Borel}}^{(0),{\rm CIPT}} (s_0)$ in the presence of IR renormalon cuts. This difference can be computed analytically for any concrete expression of the Borel function $B(u)$ and is dubbed the {\it asymptotic separation} (AS). 
The asymptotic separation describes the observed discrepancy between the CIPT and FOPT expansions for spectral function moments very well, as is shown for example for the kinematic moment in the left panel in Fig.~\ref{Plot-2} as a function the order. Note that the results shown in Fig.~\ref{Plot-2} up to order 5 (i.e.\ 6-loops for the Adler function) correspond to the QCD corrections used in state-of-the-art phenomenological analyses,\footnote{This includes the concretely known 5-loop corrections up to $c_{4,1}$~\cite{Baikov:2008jh} and an estimate for the 6-loop coefficient $c_{5,1}$.} while the orders beyond are based on a Borel model for $B[\hat D](u)$ following Ref.~\cite{Beneke:2008ad}. 
The FOPT Borel sum (red horizontal line) and the CIPT Borel sum (blue horizontal line), which is the sum of the FOPT Borel sum and the asymptotic separation, are computed from the Borel model as well.
The Borel model provides a good approximation for the true Borel function~\cite{Beneke:2012vb} and relies on the proposition that the asymptotic behavior of the gluon condensate (GC) renormalon has a sizeable contribution to the known QCD corrections up to 5-loops and therefore contains a GC contribution with a sizeable norm $N_g$. The known QCD corrections to the Adler function are fully consistent with this proposition. In this context the asymptotic separation is numerically by far dominated by the contribution from the GC renormalon.

One can also construct conceptual toy Borel models for $B[\hat D](u)$ containing only the GC renormalon, which in the context of the standard form of the OPE and the canonical renormalon calculus formalism must lead to convergent spectral function moments for weight functions such as $W_{\rm kin}$ (because for these spectral function moments all OPE corrections vanish identically). It was shown in Ref.~\cite{Hoang:2020mkw,Hoang:2021nlz} that for these moments $\delta^{(0)}_{W}(s_0)$ is indeed convergent in the FOPT expansion, while the CIPT expansion series is divergent.
This implies that the OPE corrections that need to be added to $\delta^{(0)}_W$ in the CIPT approach differ from those in the FOPT approach and, more importantly, do not have the standard form assumed in the literature. 
In other words, the CIPT expansion and the CIPT Borel representations given in Eq.~(\ref{eq:BorelCIPT}) are {\it not consistent with the standard form of the OPE as given in Eq.~(\ref{eq:OPEstandard})}.

From these results one can conclude that the consistency of the CIPT expansion can be reconciled (at least for the most part) if the GC renormalon is removed from the Adler function. 
This can be achieved through a change of scheme for the GC matrix element, which is in close analogy to the well-known use of short-distance mass renormalization schemes in favor of the pole mass for quark mass-sensitive observables~\cite{Hoang:2020iah,Beneke:2021lkq}.

\section{Renormalon-Free Gluon Condensate Scheme}

In Ref.~\cite{Benitez-Rathgeb:2022yqb} a suitable renormalon-free (RF) GC matrix element scheme was constructed in two steps. Calling the original GC matrix element $\langle \bar{G}^2 \rangle$ (which still contains the ${\cal O}(\Lambda_{\rm QCD}^4)$ renormalon) ${\overline{\rm MS}}$ GC scheme, one can first define an $R$-scale dependent renormalon-free GC matrix element $\langle G^2 \rangle (R^2)$ by the relation
\begin{align}
\label{eq: RF GC}
\langle \bar{G}^2 \rangle \equiv \langle G^2 \rangle (R^2) - R^4 \sum_{\ell=1} N_g ~ r_{\ell}^{(4,0)} \bar a^{\ell}(R^2)\,.
\end{align}
The term $N_g$ refers to the universal norm of the gluon condensate and $R$ takes the role of an IR factorization scale. The coefficients $r_{\ell}^{(4,0)}$ contain the divergent asymptotic series related to a pure  ${\cal O}(\Lambda_{\rm QCD}^4)$ renormalon and, using the $C$-scheme for the strong coupling~\cite{Boito:2016pwf} (indicated by the bar), can be given in closed form, $r_{\ell}^{(4,0)}=(\frac{1}{2})^{\ell+4 \hat b_1}\,\frac{\Gamma(\ell+4 \hat b_1)}{\Gamma(1+4 \hat b_1)}$. Numerically the $C$-scheme for the strong coupling is very close to the $\overline {\rm MS}$ scheme, see the appendix of Ref.~\cite{Benitez-Rathgeb:2022yqb} for details.   
However, the definition of Eq.~(\ref{eq: RF GC}) is somewhat inconvenient for actual phenomenological applications, since the dependence on $R$, which cancels between the subtraction series and $\langle G^2 \rangle (R^2)$, would generate a sizeable intermediate $R$-dependence. It is therefore suitable to define, in a second step a scale-invariant GC matrix element $\langle G^2\rangle^{\rm RF}$, called the RF GC scheme, by the relation
\begin{eqnarray}
	\label{eq:GCIRsubtracted2}
	\langle G^2\rangle(R^2)
	& \equiv &
	\langle G^2\rangle^{\rm RF} + N_g \, \bar c_0(R^2)\,.
\end{eqnarray}
The function $\bar c_0(R^2)$ satisfies the same $R$-evolution equation as the subtraction series on the RHS in Eq.~(\ref{eq: RF GC}) and thus of $\langle G^2\rangle(R^2)$. Because the subtraction series contains a pure ${\cal O}(\Lambda_{\rm QCD}^4)$ renormalon this $R$-evolution equation is a convergent series~\cite{Hoang:2009yr} and can even be given in closed form~\cite{Benitez-Rathgeb:2022yqb}.
A suitable choice is ($\hat b_1=\beta_1/2\beta_0^2$)
\begin{equation}
	\label{eq:subtractclosed}
	\bar c_0(R^2)  \equiv  
	- \frac{R^4\,e^{-\frac{2}{\bar a(R^2)}}}{(\bar a(R^2))^{4\hat b_1}}\,
	{\rm Re}\left[\, e^{4\pi \hat b_1 i}\, \Gamma\Big(\!-4\hat b_1,-\frac{2}{\bar a(R^2)}\,\Big)\,\right]\,,
\end{equation}
which is the FOPT Borel sum of the subtraction series itself using the common PV prescription.

The perturbative Adler function in the RF GC scheme then has the form
\begin{eqnarray}
	\label{eq:invBorelDR}
	\hat D^{\rm RF}(s,R^2) & = &
	\frac{1}{s^2}\,\Big[ 1 + \bar c_{4,0}^{(1)} \bar a(-s) \Big]\,\frac{2\pi^2}{3}\,N_g\, \bar c_0(R^2)
	\, + \,
	\sum_{\ell=1}^\infty \, \bar c_{\ell} \,\bar a^\ell(-s)\,  \qquad
	\\
	\lefteqn{
		\, - \,
		\Big[ 1 + \bar c_{4,0}^{(1)}\, \bar a(-s)  \Big]\, \frac{2\pi^2}{3}\,N_g\,\frac{R^4}{s^2}
		\sum\limits_{\ell=1}^\infty
		\Big(\frac{1}{2}\Big)^{\ell+4 \hat b_1}\,\frac{\Gamma(\ell+4 \hat b_1)}{\Gamma(1+4 \hat b_1)} \,\bar a^\ell(R^2)\,,
	}\nonumber
\end{eqnarray}
where the term involving $\bar c_0(R^2)$ must be treated strictly as a tree-level term (i.e.\ not being reexpanded again and numerically evaluated in the $C$-scheme for the strong coupling).
For the perturbation series, either using the CIPT or the FOPT renormalization scale setting prescription, {\it it is mandatory to consistently expand and truncate the other terms in $a^\ell(-s)$ or $a^\ell(s_0)$, i.e.\  using the strong coupling at a common renormalization scale}.\footnote{This expansion may be carried out using the usual $\overline{\rm MS}$ strong coupling scheme. We have done so in our numerical analyses below.} This ensures the systematic removal of the GC renormalon from the Adler function. The GC OPE correction in the RF GC scheme adopts the standard form
\begin{eqnarray}
	\label{eq:AdlerOPEGCBS}
	\delta D^{\rm OPE,RF}_{4,0}(s) & = &
	\frac{1}{s^2}\frac{2\pi^2}{3}\,\Big[ 1 -\frac{22}{81} \, \bar a(-s) \Big]\, \langle G^2\rangle^{\rm RF} \,.
\end{eqnarray}
A graphical illustration of the impact of the subtraction of the gluon condensate renormalon on the perturbative series in the RF GC scheme is given by the right panel in Fig.~\ref{Plot-2}  for the case of the kinematic moment, for $R=0.8\,m_\tau$ and considering again the Borel model for the orders beyond the 6-loop level. The Borel model provides the concrete value $N_g=0.64$ for the GC renormalon norm. 
In the RF GC scheme the CIPT and FOPT series are perfectly compatible and both approach the FOPT Borel sum given by the FOPT Borel representation in Eq. (\ref{eq:BorelFOPT}).

\begin{figure}
	\begin{center}
		\includegraphics[width=\textwidth]{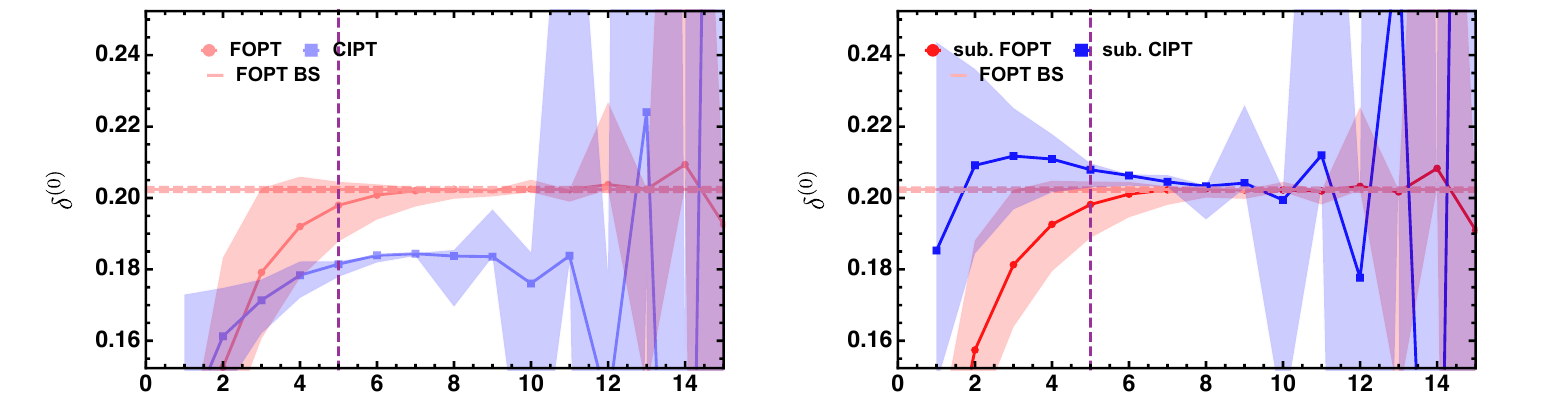}
		\caption{\label{Plot-scale-err}
			FOPT and CIPT expansion series in the  $\overline{\rm MS}$ GC scheme (left panel) as well as in the RF GC scheme for $R=0.8\,m_\tau$ and $N_g=0.64$ (right panel) for the case of the kinematic moment with $\alpha_s(m_\tau^2)=0.315$ and including renormalization scale variations. The vertical line indicates the model dependence starting at order six. The space between the dashed lines corresponds to the ambiguity of the FOPT Borel sum. Figures taken from Ref.~\cite{Benitez-Rathgeb:2022yqb}.}
	\end{center}
\end{figure}

In Fig.~\ref{Plot-scale-err} the results of Fig.~\ref{Plot-2} are shown once more, but this time we include the error bands arising from renormalization scale variations. The latter are determined by expanding in powers of $\bar a^\ell(-\xi s)$ (for CIPT) and $\bar a^\ell(\xi s_0)$ (for FOPT) and using variations in the interval $1/2\le \xi \le 2$. While the scale variation bands do not overlap at the six-loop level in the $\overline{\rm MS}$ GC scheme, they do so in the RF GC scheme. 

\section{Strong coupling predictions in the RF GC scheme}
\label{sec:strongcoupl}

The CIPT-FOPT discrepancy has been one of the dominant sources of uncertainty in $\alpha_s$ determinations from experimental data on the inclusive hadronic $\tau$ decay spectra. 
We can study the impact the RF GC can have on strong coupling determinations from data through the following simplistic toy analysis: We take the spectral function moment FOPT Borel sum of Eq.~(\ref{eq:BorelFOPT}) based on the Borel model mentioned above for the input value $\alpha_s(m_\tau^2)=0.315$ as a proxy for the perturbative part of the “experimental" moment values and determine the strong coupling from the 6-loop series (as it is common in current state-of-the-art strong coupling determinations). To estimate the theoretical uncertainty we account for renormalization scale variations (using the prescription of Fig.~\ref{Plot-scale-err}), an uncertainty in the estimate of the 6-loop coefficient $c_{5,1}$ and variations for $R$ in the range $0.7 m_\tau\le R\le 0.9 m_\tau$ (see Ref.~\cite{Benitez-Rathgeb:2022yqb} for details). We consider the spectral function moments for the  weight functions $W_{\rm kin}(x)$, $W_2(x)=(1-x)^3=1-3x+3x^2-x^3$ and $W_3(x)=2(1-x)$. For the spectral function moments associated to the weight functions $W_{\rm kin}(x)$ and $W_3(x)$ the GC OPE correction is highly suppressed (because the weight functions do not contain a quadratic term $x^2$). Spectral function moments of this kind have been the basis of all recent high precision determinations of $\alpha_s$ from hadronic $\tau$ decay data. On the other hand, for the spectral function moment associated to the weight function $W_2(x)$ the GC OPE correction unsuppressed (because $W_2$ contains a quadratic term $x^2$).

\begin{figure}
	\begin{center}
		\includegraphics[width=\textwidth]{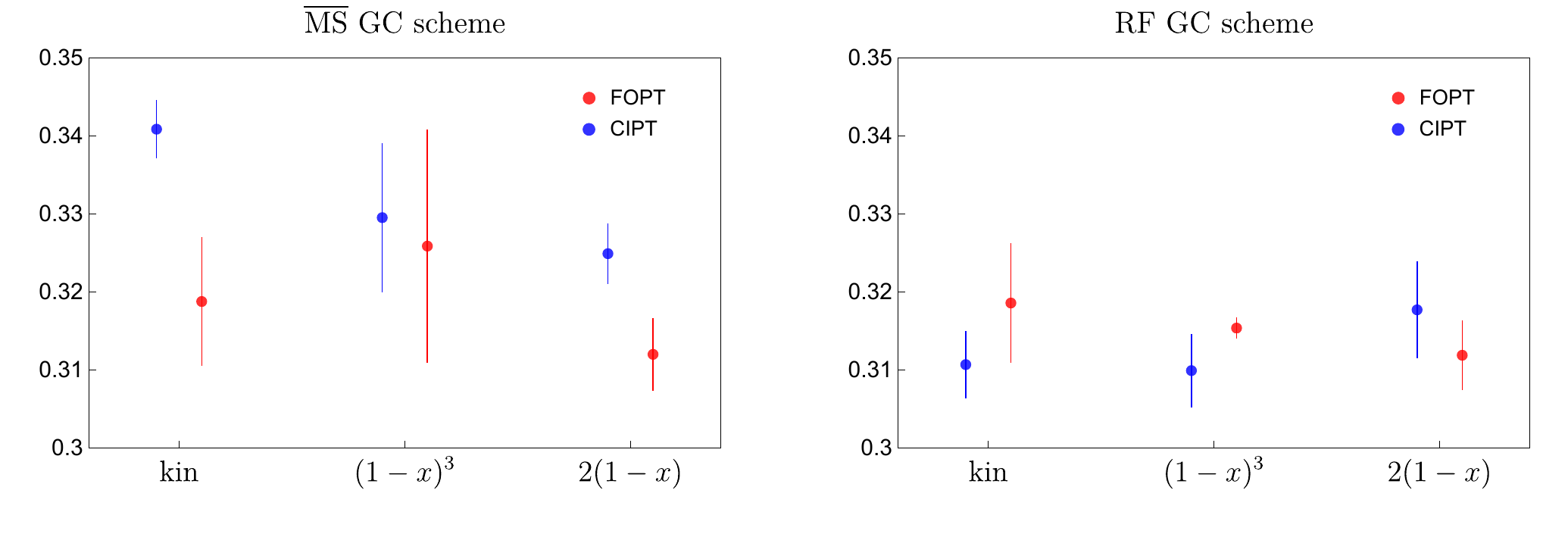}
		\caption{\label{Plot-3}
			Strong coupling results determined from the toy analysis described in Sec.~\ref{sec:strongcoupl}. The input value for the ‘experimental’ moments is $\alpha_s(m_{\tau}^2)=0.315$. Figures taken from Ref.~\cite{Benitez-Rathgeb:2022yqb}.}
	\end{center}
\end{figure}

The results are shown in Fig.~\ref{Plot-3}. The left panel shows to the resulting extracted values for $\alpha_s(m_\tau^2)$ using perturbation theory in the unsubtracted $\overline{\rm MS}$ GC scheme. 
For the kinematic moment and the one associated to the weight function $W_3$ we see the well-known CIPT-FOPT discrepancy, where the values for $\alpha_s(m_\tau^2)$ obtained from CIPT and FOPT are not compatible and where the result for  $\alpha_s(m_\tau^2)$ from the CIPT expansion comes out way too large. For the moment associated to the weight function $W_2$ the results are compatible, but the errors are very large as well, reflecting the large perturbative uncertainties for moments where the GC OPE corrections is unsuppressed.
The right panel shows the results obtained using the RF GC scheme. The CIPT and FOPT results for $\alpha_s(m_\tau^2)$ obtained from the moments for $W_{\rm kin}$ and $W_3$ are now fully compatible and consistent with the input value $\alpha_s(m_\tau^2)=0.315$. This has been achieved due to the significant modification of the CIPT expansion when the RF GC scheme is employed. Furthermore, the large uncertainties for $\alpha_s(m_\tau^2)$ obtained from the moment associated to the weight function $W_2$, which were large for CIPT as well as FOPT in the $\overline{\rm MS}$ GC scheme, have decreased significantly.  
The results show that in the RF GC scheme we can achieve two improvements. First, the CIPT-FOPT discrepancy in strong coupling determinations can be avoided and, second, high-precision analyses can be carried out with spectral function moments where the GC OPE corrections is unsuppressed. 

\section{Conclusions}

In the recent work of Refs.~\cite{Hoang:2020mkw,Hoang:2021nlz} it was shown that the Borel representations for the FOPT and CIPT hadronic $\tau$ spectral function moments are intrinsically different. In the presence of IR renormalons the different analytic properties inherent to the FOPT and CIPT expansions results in a systematic and computable disparity in their respective Borel sums, which we called the {\it asymptotic separation}. From dedicated studies of toy Borel models for the Adler function it was shown that the {\it CIPT expansion is not compatible with the standard form of the operator product expansion}. Given that the major contribution to the CIPT-FOPT discrepancy is related to the leading gluon condensate renormalon, we have shown in Ref.~\cite{Benitez-Rathgeb:2022yqb} that the discrepancy between the two approaches can be reconciled in an IR subtracted perturbation theory. In particular, we defined a new renormalon-free gluon condensate scheme \cite{Benitez-Rathgeb:2022yqb} and observed that in this scheme the perturbative series indeed converge to the same Borel sum given by the FOPT Borel representation. By considering a toy model extraction of the strong coupling, we showed that the RF GC scheme will lead to strong coupling determinations from hadronic $\tau$ decays where the CIPT-FOPT discrepancy does not arise any more. The results shown in this talk were based on the concrete value $N_g=0.64$ for gluon condensate renormalon norm which was obtained from a concrete realistic Borel model for the Adler function. In practice $N_g$ can only be determined approximately.
In Ref.~\cite{Benitez-Rathgeb:2022hfj} we have carried out complete phenomenological strong coupling determinations in the RF GC scheme, accounting also for the uncertainty in the GC renormalon norm $N_g$, which turn out to give a relatively small contribution to the final error in $\alpha_s(m_\tau^2)$.


\end{document}